\documentclass[12pt]{article}
\usepackage{amssymb,amsmath,slashed}
\include{epsf}

\newcommand{\e}{{\rm e}}
\newcommand{\ii}{{\rm i}}
\newcommand{\dd}{{\rm d}}
\newcommand{\eqn}[1]{(\ref{#1})}
\def\appendix#1{\addtocounter{section}{1}\setcounter{equation}{0}
\renewcommand{\thesection}{\Alph{section}}
\section*{
\thesection\protect\indent \parbox[t]{11.715cm} {#1}}
\addcontentsline{toc}{section}{Appendix\thesection\ \ \ #1} }
\newcommand{\complex}{{\mathbb C}} 

\def\ket#1{\left| #1\right\rangle}

\def\ketbra#1#2{\left| #1\right\rangle\left\langle #2\right|}

\newcommand{\tr}[1]{\:{\rm tr}\,#1}
\newcommand{\Tr}[1]{\:{\rm Tr}\,#1}
\def\one{\mbox{1 \kern-.59em {\rm l}}}
\newcommand{\be}{\begin{equation}}
\newcommand{\ee}{\end{equation}}
\newcommand{\beq}{\begin{equation}}
\newcommand{\eeq}{\end{equation}}
\newcommand{\bea}{\begin{eqnarray}}
\newcommand{\eea}{\end{eqnarray}}

\newcommand{\del}{\partial}

\declareslashed{}{/}{.1}{0}{D}

\begin{document}
\begin{titlepage}
\begin{flushright}

\baselineskip=12pt DSF/1/2010\\ SPbU-IP-10-01 \\ICCUB-10-004\\
\hfill{ }\\
\end{flushright}

\begin{center}

\baselineskip=24pt

{\Large\bf Bosonic Spectral Action Induced from Anomaly Cancelation
}

\baselineskip=14pt

\vspace{1cm}

{\bf A.A.\ Andrianov$^{1,2}$ and Fedele Lizzi$^{1,3}$}
\\[6mm]
 $^{1}${\it High Energy Physics Group, Dept. Estructura i Constituents
de la Mat\`eria, \\Universitat de Barcelona, Diagonal 647, 08028
Barcelona, Catalonia, Spain \\and
Institut de Ci\`encies del Cosmos, UB, Barcelona}\\
 $^{2}${\it V.A. Fock Department of Theoretical Physics, Sankt-Petersburg State University,
198504 St. Petersburg, Russia}\\
$^{3}${\it Dipartimento di Scienze Fisiche, Universit\`{a} di
Napoli {\sl Federico II}}\\ and {\it INFN, Sezione di Napoli}\\
{\it Monte S.~Angelo, Via Cintia, 80126 Napoli, Italy}\\{\small\tt
andrianov@bo.infn.it, fedele.lizzi@na.infn.it }
\\[10mm]

\end{center}

\vskip 2 cm

\begin{abstract}
We show how (a slight modification of) the noncommutative geometry
bosonic spectral action can be obtained by the cancelation of the
scale anomaly of the fermionic action. In this sense the standard
model coupled with gravity is induced by the quantum nature of the
fermions. The regularization used is very natural in noncommutative
geometry and puts the bosonic and fermionic action on a similar
footing.
\end{abstract}

\end{titlepage}

\section{Introduction}

Classical general relativity is a \emph{geometrical} theory describing how the
curvature of spacetime influences the motion of classical bodies. The standard
model of elementary particles is on the other side a quantum field theory and
the difficulties in reconciling the two are well known. The noncommutative
geometry programme~\cite{Connesbook,ConnesMarcollibook} aims at a
generalization of ordinary geometry along the lines of the one made to describe
quantum mechanically the phase space. The programme is based on a transcription
of ordinary (commutative) geometry in algebraic terms, based on the duality
between commutative $C^*$-algebras and topological spaces. The setting is then
generalized to noncommutative algebras which may or may not be matrix algebras
over an ordinary space. In the former case one talks of \emph{almost
commutative spaces}. The geometrical information on the space is given by the
spectral data defined by the \emph{spectral triple}, comprised of an
$*$-algebra $\mathcal A$, a fermionic Hilbert space on which the algebra is
represented by bounded operators and a Hermitian (generalized) Dirac operator
$D_0$. Geometry is then translated into the spectral properties of these
operators. All usual concepts obtain an algebraic equivalent: integrals
generalize to traces, differential forms are operators obtained commuting
functions with the Dirac operator, and a dictionary translating the concept
ordinary geometry in this language is being built. The setting is solid and it
generalizes naturally to the case in which the algebra is noncommutative (hence
the name of noncommutative geometry), and an underlying point geometry may not
exist. Details can be found in~\cite{Connesbook, Landibook, Ticos, Madore},
which by now are classic descriptions of noncommutative geometry.

Already at the classical level the construction requires the
presence of fermions. While it is still impossible to ``hear the
shape of a fermionic drum'' because of isospectral
manifolds~\cite{Baer}, the Dirac operator carries more information
than the Laplacian~\cite{Connesreconstruct}. This gives a centrality
to fermions in geometry. The elements of the algebra of functions on
a manifold are ``bosonic'', and they capture only the topology of
the space (via continuity of the functions). The full geometrical
information requires necessarily the presence of fermions and their
Hilbert space on which the Dirac operator is defined.

Connes' approach to the standard model is the attempt to understand
which kind of (noncommutative) geometry gives rise to the standard
model of elementary particles coupled with gravity. The most
complete formulation of this is given by the \emph{spectral action},
described in the next section. The starting point is an almost
commutative geometry product of the algebra of functions on ordinary
spacetime times a finite dimensional matrix algebra. It comprises of
a bosonic and a fermionic part, which are treated somewhat
differently, and it reproduces the Lagrangian of the fermions of the
standard model in a curved background, and contains all required
terms for the Higgs mechanism of symmetry breaking. Its input are
the masses (and mixings) of the fermions, and the coupling constants
at low energy. The action must be read in the Wilsonian sense and
undergoes renormalization, which is done in the usual way. The
result is the full action of the standard model coupled with
gravity, with some extra phenomenologically relevant terms. The mass
of the Higgs can be calculated in terms of the other parameters of
the theory, and while its value (170~GeV) may have been recently
excluded by {\sl Fermilab} data, it is still surprising that a
purely geometrical theory is capable to make specific predictions
which are of the correct order of magnitude.

The purpose of this paper is to show that the bosonic spectral
action is a consequence of the fermionic action and the cancelation
of the scale anomaly. We see that the spectral action is a
\emph{quantum} effect of the fermionic action, and its
regularization. A fact already noted in a different context
in~\cite{AndrianovBonoraGamboa, AndrianovBonora1, AndrianovBonora2}.
The crucial aspect is the cancelation of the anomaly which develops
under a \emph{spectral} regularization of the fermionic action. Our
calculation is totally general and comes prior to the application to
the standard model. We will therefore be very general in our
treatment of the action and comment on the standard model where
appropriate. In a sense we explicitly show that the spectral action
is induced perturbatively by the action for matter, which is the old
idea of Sakharov~\cite{Sakharov} (for a modern review
see~\cite{Visser}). In fact it has been already shown by
Yu.~Novozhilov and D.~Vassilevich~\cite{NovozhilovVassilevich} that
this anomaly induces quantum gravity.

In section 2 we briefly introduce the spectral action and discuss
its properties under scale invariance. In section 3 we discuss
anomalies in the present context. In section 4 we show with an
explicit calculation how a slightly modified version of the bosonic
spectral action is the term required to cancel the scale anomaly,
and in the following section we show explicitly the slight
modifications, which amount to a change of some coefficients. A
final section contains some final remarks.

\section{Spectral Physics and Scale Invariance}
\setcounter{equation}{0}

The point of view that we will take here is that the main
characteristics of the standard model coupled with gravity can be
obtained from the study of the spectral properties of a suitable
algebra of functions on spacetime (the fields) and a generalized
Dirac operator. We will first review the main aspects of the
spectral action, stressing the differences between the fermionic and
bosonic parts, and then discuss scale invariance in this context.

\subsection{The Spectral Action}

The spectral action, in the spirit of noncommutative geometry, depends on the
spectral data of the space, defined by a spectral triple. In the description of
the standard model of~\cite{ConnesLott, Connesreal, ChamseddineConnes, AC2M2}
the space is the tensor product of ordinary (Euclidean) spacetime by an inner
space described by a finite dimensional matrix algebra. The algebra of this
extended spacetime acts as operators on an Hilbert space which comprises the
fermions\footnote{A projection may be necessary to avoid fermion
doubling~\cite{LMMS,GraciaBondiaSchukerIochum,AC2M2} but this is not essential
at this stage. Likewise the real structure $J$ and the chirality $\gamma$,
which are otherwise crucial~\cite{Connesreal}, play no role in this
discussion.}. The metric properties of the space, as well as the differential
structure and the action, depend on the operator $D_0$. This operator
``fluctuates'' with the addition of a Hermitian one-form which we will
generically indicate with $A$ and that can be expressed in terms of the algebra
of functions which defines spacetime:
\bea
D&=&D_0+A\nonumber\\
A&=&\sum_i a_i [D_0,b_i]
\eea
with $a_i,b_i\in\mathcal A$.
In the case of the standard model plus gravity the connection $A$ comprises both the Levi-Civita and the gauge connections. In this case the geometry is the product of the continuous (commutative) spacetime times a noncommutative inner space described by a finite dimensional algebra, i.e.:
\bea
\mathcal A &=& C(M) \otimes \mathcal A_F\nonumber\\
 \mathcal A_F&=&\complex\oplus\mathbb H\oplus
M_3(\complex) \label{algebraSM}
\eea
where $C(M)$ is the algebra of continuous functions on spacetime
$M$, $\mathbb H$ is the algebra of quaternions (whose unitary
subgroup is $SU(2)$) and $M_3(\complex)$ is the algebra of $3\times
3$ complex valued matrices. The unimodular (unitary and unit
determinant) elements of $\mathcal A$ form the standard model group
$U(1)\times SU(2) \times SU(3)$. This algebra is represented on the
Hilbert space
\be
\mathcal H=L_2(sp(M))\otimes \mathcal H_F \label{hilbertSM}
\ee
the tensor product of
spinors on $M$, times a finite dimensional space which comprises all
fermions, in three generations. Also the Dirac operator has two
parts
\be
D_0=D_M\otimes\one+\gamma_5\otimes D_F \label{diracSM}
\ee
where $D_M$ is the ordinary Dirac operator on $M$, and $D_F$ is a
finite matrix which carries the information of the values of the
masses of the fermions and their Cabibbo mixings (including that of
neutrinos).

Although the successes of the spectral action are obviously related
to the standard model, in the following we will be more general, and
our considerations will be valid for any spectral triple with a
representation on the fermionic Hilbert. Given the ingredients of
the triple, the spectral action comprises of two parts, one bosonic
and one fermionic
\be
S=S_B+S_F= \Tr \chi\left(\frac{D^2}{\Lambda^2}\right) + \langle\psi|
D \psi\rangle \label{action}
\ee
where $\Tr$ is the usual operator trace, $\Lambda$ is the energy cutoff of renormalization and $\chi$ is a positive function. Its particular shape is not essential as long as $\chi(0)=1$ and $\chi(x)=0$ for $x \gtrsim 1$.
The fermionic part of the action is the usual integral over spacetime of the expectation value of the Dirac operator.
The bosonic action contains the renormalization cutoff in its very definition, and therefore it must be considered in the Wilson renormalization scheme. On the other side the fermionic part is in general divergent and it must be renormalized as well.

The bosonic spectral action is a sum of residues \cite{Gilkey} and
can be expanded in a power series in terms of $\Lambda^{-1}$ as
\be
S_B=\sum_n f_n\, a_n(D^2/\Lambda^2)
\ee
where the $f_n$ are the momenta of $\chi$
\bea
f_0&=&\int_0^\infty \dd x\, x  \chi(x)\nonumber\\
f_2&=&\int_0^\infty \dd x\,   \chi(x)\nonumber\\
f_{2n+4}&=&(-1)^n \del^n_x \chi(x)\bigg|_{x=0} \ \ n\geq 0
\label{fcoeff}
\eea
while the $a_n$ are the Seeley-de Witt coefficients \cite{Gilkey}
which in this case vanish for $n$ odd. We now give the form of the
first three $a$'s as functions of the terms of the square of the
Dirac operator, using essentially the notations
of~\cite{ChamseddineConnesscale} (see
also~\cite{Vassilevich:2003xt}). Consider a $D^2$ of the form
\be
D^2=g^{\mu\nu}\del_\mu\del_\nu\one+\alpha^\mu\del_\mu+\beta
\ee
then define
\bea
\omega_\mu&=&\frac12 g_{\mu\nu}\left(\alpha^\nu+g^{\sigma\rho} \Gamma^\nu_{\sigma\rho}\one\right)\nonumber\\
\Omega_{\mu\nu}&=&\del_\mu\omega_\nu-\del_\nu\omega_\mu+[\omega_\mu,\omega_\nu]\nonumber\\
E&=&\beta-g^{\mu\nu}\left(\del_\mu\omega_\nu+\omega_\mu\omega_\nu-\Gamma^\rho_{\mu\nu}\omega_\rho\right)
\label{laplawithoutdilaton}
\eea
then
\bea
a_0&=&\frac{\Lambda^4}{16\pi^2}\int\dd x^4 \sqrt{g}
\tr\one_F\nonumber\\
a_2&=&\frac{\Lambda^2}{16\pi^2}\int\dd x^4 \sqrt{g}
\tr\left(-\frac R6+E\right)\nonumber\\
a_4&=&\frac{1}{16\pi^2}\frac{1}{360}\int\dd x^4 \sqrt{g}
\tr(-12\nabla^\mu\nabla_\mu R +5R^2-2R_{\mu\nu}R^{\mu\nu}\nonumber\\
&&+2R_{\mu\nu\sigma\rho}R^{\mu\nu\sigma\rho}-60RE+180E^2+60\nabla^\mu\nabla_\mu
E+30\Omega_{\mu\nu}\Omega^{\mu\nu}) \label{spectralcoeff}
\eea
where by $\tr$ we indicate the trace over the inner indices of the
finite algebra $\mathcal A_F$.

The action with the spectral triple described by the
data~\eqn{algebraSM}-\eqn{diracSM} reproduces correctly~\cite{AC2M2}
the standard model coupled with gravity and it has predictive power
in relation to the Higgs mass for example, and it has been applied
recently to cosmology as well~\cite{NelsonSakellariadu,
MarcolliPierpaoli}. We refrain to write in full glory all of the
terms of the action which takes a full page and can be found
in~\cite[Sect.~4.1]{AC2M2}.

However the action is basically a classical quantity, and the
renormalization is performed, especially in the fermionic sector,
using standard field theory techniques. The model has the three
coupling constants equal at the renormalization point, as is the case
of $SU(5)$ non-supersymmetric unification, and hence some of the
predictions are similar to the ones of the this theory.

\subsection{The different nature of the Bosonic and Fermionic Actions}

As they stand the bosonic and the fermionic parts of the
action~\eqn{action} are very different. The bosonic one is always
finite and it depends on the cutoff $\Lambda$. It is the usual trace
of an operator and it does not diverge because of the presence of
the function $\chi$ which regularizes. In the case of $\chi$ being
the characteristic function of the interval, i.e.
\be
\chi(x)=\left\{\begin{array}{cc}0~ & x<0\\ 1~ & x\in[0,1]\\ 0~ & x>1
\end{array}\right. \label{sharpcutoff}
\ee
or a smooth version of it, the bosonic spectral action simply  counts
the eigenvalues of $D$ which are less than the cutoff $\Lambda$.

The fermionic action on the contrary is divergent, and will require
renormalization. It is formulated as an usual integral, which in
this context (in four dimensions) is the Dixmier trace:
\be
\int \dd x f=\Tr_\omega |D|^{-4} f
\ee
where the Dixmier trace of an operator $O$ with eigenvalues $o_n$
(ordered in decreasing order, repeated in case of degeneracy) is:
\be
\Tr_\omega O=\lim_{N\to\infty}\frac{1}{\log N} \sum_{n=0}^No_n
\ee
The integral/Dixmier trace has to be regularized. Since the action
is written as an usual integral, the renormalization analysis can be
done in a variety of ways. In this process however some quantum
symmetries can be lost, and the theory can develop an anomaly. In
Sect.~\ref{se:regferm} we will use a regularization which is rather
similar to the one used for the fermionic part. It remains the fact
that the different treatment of the two parts of the action seems
\emph{ad hoc}, and it would be desirable to have them to be part of
a more uniform approach.

\subsection{Scale Invariance in the Spectral Action}

The standard model is classically invariant against scale transformations if one
ignores the mass terms, which can be done at high energy. The lack of full
invariance can also be compensated by the introduction of a~\emph{dilaton}
field, or by giving nontrivial transformation properties to the masses under a
scale transformations. In this paper will only discuss the case of a global rescaling by
a constant parameter.

We want to define our theory to be invariant under rescaling defined as
\bea
x^\mu&\to&\e^\phi x^\mu\nonumber\\
\psi&\to& \e^{-\frac32\phi} \psi\nonumber\\
D&\to& \e^{-\frac12\phi}D\e^{-\frac12\phi} \label{scaleinvariance}
\eea
where for the scope of this paper $\e^\phi$ is a constant real
parameter. In future we hope to discuss the case of $\phi$ being a
(dilaton) field.

Note that since the rescaling involves also the matrix part of $D$,
we must also rescale the masses of the fermions. This is tantamount
to a change of the unit of measurement and, in the absence of a
dimensional scale, is an exact symmetry of the classical theory.
This classical symmetry can however develop an anomaly, namely not be
a symmetry of the (renormalized) quantum theory anymore. In the next
section we will discuss the presence of an anomaly due to the
breaking of this symmetry.

\section{Anomalies}
\setcounter{equation}{0}

In the present context we have an anomaly: a classical theory is
invariant for a symmetry, but the quantum theory, due to unavoidable
regularization, does not possess this symmetry anymore. If also
the quantum theory is required to be symmetric then the symmetry can
be restored by the addition of extra terms in the action. A
textbook introduction to anomalies can be found in
\cite{Fujikawabook}.

As explained in the previous chapter, the notion of scale anomaly is
attached to the dilatation  of both coordinates,  fields and
mass-like parameters according to their dimensionalities,
Eq.~\eqn{scaleinvariance}. Evidently, in the absence of UV
divergences, there is no scale anomaly which therefore can be
correlated to rescaling of a cutoff in the theory. In general the
dilatation need not be constant, and the quantum field corresponding
is called the \emph{dilaton}.

There is the adjacent notion of Weyl or conformal anomaly, which is
closely related. It is based on the symmetry against local Weyl
dilatation of the metric accompanied by an appropriate
transformation of the dilaton field dressing all mass-like vertices
in order to make homogeneous the entire transformation of the Dirac
Lagrangian, that is
\bea
g^{\mu\nu}&\to&\e^{2\alpha} g^{\mu\nu}\nonumber\\
\psi&\to& \e^{-\frac32\alpha} \psi\nonumber\\
D&\to& \e^{-\frac12\alpha}D\e^{-\frac12\alpha}
\label{weylinvariance}
\eea
while $x^\mu$ is left untouched and in this case $\alpha$ is local
function of $x$. Scale and Weyl anomalies are closely and directly
related.

In the functional integral the proper measure to use is the sum over
all configurations of $\tilde\psi=(-g)^{1/4}\psi$ and
$\tilde{\bar\psi}=(-g)^{1/4}\bar\psi$, which we will indicate as
$[\dd\psi][\dd\bar\psi]$, the partition function (which we define
below) is formally invariant for the scale (or Weyl) transformation,
but the regularization procedure spoils this formal invariance,
giving rise to the anomaly.

In spite of the fact that the  generator of Weyl dilatation is
localized, the transformation of quantum action reveals an anomalous
breaking of the symmetry (see the history in~\cite{Duff}). The
reason is that the Dirac operator is unbounded whereas any local
transformation of fields and/or operators is singular as an integral
operator, and they don't commute.  When calculating the determinant
of the product of Dirac operator and its local Weyl transformation
one cannot just factorize it to prove the essential invariance of
the fermionic quantum action, first one has to make the product
finite and therefore perform a regularization. As the above
mentioned operators don't commute their regularization  may entail
non-factorizability - a non-commutative residue~\cite{wodz}, which
can be interpreted  as a conformal non-invariance of the measure in
the path integral approach \cite{fuji}. Symbolically one can present
the anomalous action for fermions as,
\be
||\e^{-\frac12\phi}D\e^{-\frac12\phi}||_{Reg} =
||\e^{-\phi}||_{Reg}\times  ||D||_{Reg} \times
\exp(-S_{anom}(\mbox{external fields})) . \label{notinvariantdet}
\ee
In the next section we will apply this procedure in the concrete
example of the spectral action.

\section{Bosonic Action from Scale Anomaly for Fermions}
\setcounter{equation}{0}

In this section, which forms the central part of the paper, we
argue that the bosonic part of the action can be seen as emerging
naturally from the regulated fermionic action as the term
necessary to compensate the scale anomaly.

Although most of discussion about the renormalization of the
spectral action~\eqn{action} has been concentrated on its bosonic
part, here we start from the fermionic action which for the purposes
of this section we write as
\be
S_\psi=\int \dd x \bar\psi D \psi \label{fermaction}
\ee
In the following we will analyze its \emph{quantum} behaviour under scale transformations.

\subsection{Regularization of the Fermionic Action \label{se:regferm}}

The action~\eqn{fermaction} appears in the partition function of the
theory:
\be
Z(D)=\int [\dd\psi] [\dd\bar\psi] e^{-S_\psi} =\det(D)\times\mbox{const} ,
\ee
where the last equality is of course just formal because the
expression is divergent and needs regularizing. The writing of the
fermionic action in this form (as a Pfaffian) is instrumental in the
solution of the fermion doubling problem \cite{LMMS, AC2M2}.

The regularization can be done in
several ways but in the spirit of noncommutative geometry and the
spectral action the most natural one is a truncation of the spectrum
of the Dirac operator. This regularization scheme has been
introduced by one of us together with L.~Bonora and R.~Gamboa-Saravi
in~\cite{AndrianovBonoraGamboa, AndrianovBonora1, AndrianovBonora2}.
The energy cutoff is enforced by considering only the first $N$
eigenvalues of $D$. Consider the projector
\be
P_N=\sum_{n=0}^N \ketbra{\lambda_n}{\lambda_n}
\ee
where $\lambda_n$ are the eigenvalues of $D$ in increasing order
(repeated according to possible multiplicities), and
$\ket{\lambda_n}$ a corresponding orthonormal basis.  The integer $N$ is a function of the cutoff and is defined as
\be
N=\max n \ \mbox{such that} \ \lambda_n\leq \Lambda
\ee
This means that we are
effectively using the $N^{\mathrm{th}}$ eigenvalue as cutoff.

We define the
regularized partition function\footnote{Although $P_N$ commutes with
$D$ we prefer to use a more symmetric notation.}
\be
Z_\Lambda(D)=\prod_{n=0}^N\lambda_n=\det\left(\one-P_N+P_N\frac{D}{\Lambda}P_N\right)
\ee
In this way we can define the fermionic action in an intrinsic way,
without reference to the Dixmier trace (integral) in a formulation
which is purely spectral.

The regularized partition function $Z_\Lambda$ has a well defined meaning.
Expressing $\psi$ and $\bar\psi$ as
\bea
\psi=\sum_{n=0}^\infty a_n\ket{\lambda_n}\nonumber\\
\bar\psi=\sum_{n=0}^\infty b_n\ket{\lambda_n}
\eea
with $a_n$ and $b_n$ anticommuting (Grassman) quantities. Then
$Z_\Lambda$ becomes (performing the integration over Grassman
variables for the last step)
\be
Z_\Lambda(D)=\int\prod_{n=0}^N \dd a_n \dd b_n \e^{-\sum_{n=0}^N b_n
\frac{\lambda_n}\Lambda a_n}=\det\left(D_N\right)
\ee
where we defined
\be
D_N=1-P_N+P_N\frac{D}{\Lambda}P_N .
\ee
In the basis in which $D/\Lambda$ is diagonal it corresponds to set to 1
all eigenvalues larger than 1. Note that $D_N$ is dimensionless and depends on $\Lambda$ both explicitly and intrinsically via the dependence of $N$ and $P_N$.

Since $P_N$ commutes with $D$. It is possible to give an explicit
functional expression to the projector in terms of the cutoff:
\be
P_N=\Theta\left(1-\frac{D^2}{\Lambda^2}\right)=\int\dd\alpha\,\frac1{2\pi\ii(\alpha- \ii\epsilon)}
\e^{\ii\alpha\big(1-\frac{D^2}{\Lambda^2}\big)}
\ee
where $\Theta$ is the Heaviside step function.

\subsection{Cancelation of the Anomaly and the Bosonic Action}

The regulated determinant is not invariant under scale
transformation, and we are in the case of \eqn{notinvariantdet}. Accordingly
the regulated partition function develops an anomaly. We have
therefore to add another term to the action which will cancel this
anomaly.

The action $S_F$ is invariant under~\eqn{scaleinvariance} but the
partition function is not, thus we need  to add another term to
the action to compensate this lack of invariance at the quantum
level. This calculation has been performed in~\cite{AANN} in the QCD
context, and applied to gravity in~\cite{NovozhilovVassilevich}.

Let us see in a very heuristic way with $\phi$ constant why the effective action $S_{\mathrm{eff}}$
is nothing but the spectral action with the function $\chi$ being a sharp
cutoff. In this case $N$ is just a number of eigenvalues smaller that
$\Lambda$, and thereby
\be
\Tr\chi\left(\frac{D^2}{\Lambda^2}\right)=\Tr P_N=N
\ee
It is worth recalling again that the integer $N$ depends on the cutoff $\Lambda$, on the Dirac operator $D$ and also on the function $\chi$ which we have chosen to be a sharp cutoff.

Then the compensating term -- the effective action, will be defined by
\be
{Z_{\mathrm{inv}}}_\Lambda(D)=Z_\Lambda(D)
\int\dd\phi\,
\e^{-S_{\mathrm{anom}}}
\ee
where the effective action will be depending on $N$ and hence the
cutoff $\Lambda$ and on $\phi$.
Define
\be
Z_{\mathrm{inv}\Lambda}(D)=\int\dd\phi Z_\Lambda(\e^{-\frac12\phi} D\e^{-\frac12\phi})
\ee
then
\be
S_{\mathrm{anom}}= \log Z^{-1}_\Lambda(D)Z_\Lambda(\e^{-\frac12\phi}
D\e^{-\frac12\phi})
\ee

Let us designate
\be
Z_t=Z_\Lambda(\e^{-\frac t2\phi} D\e^{-\frac t2\phi})
\ee
therefore $Z_0=Z_\Lambda(D)$ and
\be
{Z_{\mathrm{inv}}}_N(D)Z^{-1}_\Lambda(D)=\int\dd\phi \frac{Z_1}{Z_0}
\ee
and hence
\be
S_{\mathrm{eff}}=-\int_0^1\dd t \del_t \log Z_t =-\int_0^1\dd t
\frac{\del_t Z_t}{Z_t}
\ee
We have the following relation that can easily proven:
\be
D_N^{-1}=(1-P_N+P_NDP_N)^{-1}=1-P_N+P_ND^{-1}P_N
\ee
and
\bea
\del_t Z_t&=&\del_t\det (\e^{-\frac t2\phi} D\e^{-\frac
t2\phi})_N\nonumber\\
&=&\del_t \e^{\tr\log(1-P_N+\e^{-\frac t2\phi} D_N\e^{-\frac
t2\phi})}\nonumber\\
&=&\Tr(\del_t\log(1-P_N+\e^{-\frac t2} D_N\e^{-\frac
t2\phi})Z_t\nonumber\\
&=&-\Tr((1-P_N+\e^{-\frac t2\phi} D_N\e^{-\frac
t2\phi})^{-1}\phi\e^{-\frac t2\phi}D_N\e^{-\frac t2\phi})Z_t\nonumber\\
&=&-\phi Z_t\tr P_N
\eea
and therefore
\be
S_{\mathrm{anom}}=\int_0^1\dd t\, \phi\tr P_N \label{Sanomal}
\ee
which is indeed a structure very similar to the spectral action in the case $\chi$ as in~\eqn{sharpcutoff}.

\section{The scale invariant Spectral Action}

The calculations of the modified spectral action are very similar to
the ones for the regular spectral action and were done by
Chamseddine and Connes in \cite{ChamseddineConnesscale} for the more
general case of a $x$ dependent $\phi$. We can read the
modifications to the spectral action from their work simply setting
to zero the derivatives of $\phi$ and then carrying out the integral
in~\eqn{Sanomal}. The rescaled action with the new Dirac operator,
in this case of constant rescaling, gives just a correction of the
Seeley-de Witt coefficient of a very simple kind
\be
a_n\to a_n'=\e^{(4-n)\phi}a_n \label{arho}
\ee
while the coefficients $f_n$ in~\eqn{fcoeff} for the case of a
$\chi$ the characteristic function of the interval are:
\be
f_0=\frac12\ \ ; \ f_2= 1 \ ; \ f_4=1 \ ; f_n=0,\ n> 4
\ee
The fermionic action remains invariant.

We can now perform easily the integral in $t$ of~\eqn{Sanomal}
noting that $t$ appears always together with $\phi$, therefore with
the change of variables $t'=\phi t$ we have that
\be
S_{\mathrm{anom}}=\int_0^\phi \dd t' \sum_n \e^{(4-n)t'} a_n f_n=
\frac{1}{8} (e^{4\phi}-1) a_0 + \frac{1}{2} (e^{2\phi}-1) a_2 + \phi
a_4
\ee
A different cutoff function will give some slightly different
coefficient with the appearance of higher Seeley-de~Witt
coefficient. We see that the changes from the spectral action are
rather small, the constant $\phi$ appears in multiplicative factors.
It will play a role in the full renormalized theory where the
dependence on $\phi$ can be eliminated at the expense of the
fundamental scale $\Lambda$, given the phenomenological input of the
cosmological constant and the electroweak scale. We leave this to
another project.

\section{Final Remarks}
There are two obvious directions of development of the ideas of this
paper. On one side one can apply this to the detailed spectral
action for the standard model coupled with gravity~\cite{AC2M2}.
Since the structure of that spectral action is very similar to the
one discussed here, we expect the same coefficients to appear, but
we have not checked this. The second development is the gauging of
the symmetry, i.e.\ consider $\phi$ to be a dilaton field. This
dilaton may play an important role in the inflationary epoch and
have a role for the solution of hierarchy problem~\cite{LMMSinfla,
LMM, Chamseddinegravity}.  In this case however the field would not
commute with $D$, and in particular we would have that
$\e^{-\frac12\phi}D^2\e^{-\frac12\phi}\neq(\e^{-\frac12\phi}D\e^{-\frac12\phi})^2$
and this will change things. Moreover terms with derivatives of
$\phi$ would appear, like the kinetic term for the dilaton, and
therefore the details of the calculations will change, causing
probably changes in the coefficients of the expansion. The
conceptual framework will however remain unchanged.

We have seen how the cancelation of anomalies induces the spectral
action, and hence gravity at a quantum level. We have used only
global scale invariance (and a rescaling of the masses), in other
words the statement of the invariance of the theory is just
invariance under a change of the unit of measurement. This is
symmetry is classically exact, but the presence of a cutoff scale
spoils it. What is interesting from the point of view of
noncommutative geometry is that this scheme favors a sort of
``fermion predominance'', i.e.\ the natural fundamental fields are
the fermions, moving in a fixed background, which is fixed since the
action does not contain the terms for the self-interaction of the
gauge and gravitational degrees of freedom. But quantization, and
the ensuing anomaly, induce the spectral action, which contains the
gauge and gravitational interaction. In some sense \emph{matter was
created before light!}

\paragraph{Acknowledgments}
We both thank the University of Barcelona and the Institut de
Ciencies del Cosmos, and in particular Domenec Espriu and Joaquim
Gomis, for the hospitality which fostered our collaboration. This
work has been supported in part by CUR Generalitat de Catalunya
under project 2009SGR502. The work of A.A.A. was  supported
Grants RFBR 09-02-00073 and 09-01-12179-ofi-m.

\end{document}